\documentclass[]{aa} 

\usepackage{graphicx}
\usepackage{natbib}
\usepackage{txfonts}
\begin{document}
   \title{The wind of the M-type AGB star \object{RT\,Virginis} probed by VLTI/MIDI\thanks{Based on observations made with the Very Large Telescope Interferometer at Paranal Observatory under programs 083.D-0234 and 086.D-0737 (Open Time Observations).}}


   \author{S.~Sacuto\inst{1}, S.~Ramstedt\inst{1,2}, S.~H\"ofner\inst{1}, H.~Olofsson\inst{3}, S.~Bladh\inst{1}, K.~Eriksson\inst{1}, B.~Aringer\inst{4}, D.~Klotz\inst{4}, and M.~Maercker\inst{2}
          }

   \institute{Department of Physics \& Astronomy, Division of Astronomy \& Space Physics, Uppsala University, Box 516, 751 20 Uppsala, Sweden\\
              \email{stephane.sacuto@physics.uu.se}
         \and Argelander Institut f¨ur Astronomie, Auf dem H¨ugel 71, DE-53121 Bonn, Germany
         \and Onsala Space Observatory, Dept. of Radio and Space Science, Chalmers University of Technology, SE-43992 Onsala, Sweden
         \and University of Vienna, Department of Astrophysics, T\"urkenschanzstra\ss e 17, A-1180 Vienna, Austria
             }

   \date{Received ; accepted}
 
  \abstract 
   {}
   {We study the circumstellar environment of the M-type AGB star \object{RT\,Vir} using mid-infrared high spatial resolution observations from the ESO-VLTI focal instrument MIDI. The aim of this study is to provide observational constraints on theoretical prediction that the winds of M-type AGB objects can be driven by photon scattering on iron-free silicate grains located in the close environment (about 2 to 3 stellar radii) of the star.} 
   {We interpreted spectro-interferometric data, first using wavelength-dependent geometric models. We then used a self-consistent dynamic model atmosphere containing a time-dependent description of grain growth for pure forsterite dust particles to reproduce the photometric, spectrometric, and interferometric measurements of \object{RT\,Vir}. Since the hydrodynamic computation needs stellar parameters as input, a considerable effort was first made to determine these parameters.} 
   {MIDI differential phases reveal the presence of an asymmetry in the stellar vicinity. Results from the geometrical modeling give us clues to the presence of aluminum and silicate dust in the close circumstellar environment ($<$~5 stellar radii). Comparison between spectro-interferometric data and a self-consistent dust-driven wind model reveals that silicate dust has to be present in the region between 2 to 3 stellar radii to reproduce the 59 and 63\,m baseline visibility measurements around 9.8\,$\mu$m. This gives additional observational evidence in favor of winds driven by photon scattering on iron-free silicate grains located in the close vicinity of an M-type star. However, other sources of opacity are clearly missing to reproduce the 10-13\,$\mu$m visibility measurements for all baselines.} 
   {This study is a first attempt to understand the wind mechanism of M-type AGB stars by comparing photometric, spectrometric, and interferometric measurements with state-of-the-art, self-consistent dust-driven wind models. The agreement of the dynamic model atmosphere with interferometric measurements in the 8-10\,$\mu$m spectral region gives additional observational evidence that the winds of M-type stars can be driven by photon scattering on iron-free silicate grains. Finally, a larger statistical study and progress in advanced self-consistent 3D modeling are still required to solve the remaining problems.}

   
       \authorrunning{Sacuto et al.}          
    \titlerunning{The wind of \object{RT\,Virginis}}

   \maketitle
%

\section{Introduction}
\label{intro}

The mass loss of AGB stars is commonly ascribed to a two-step process, where the pulsation of the star creates atmospheric shock waves that push the matter to conditions where both temperature and density allow the formation of dust grains. When the dust opacity is large enough, grains receive enough momentum through radiative pressure to be accelerated away from the star and to drag the gas along by collisions, causing a slow outflow.

Although this scenario has been successfully tested for C-type AGB stars by comparing detailed models with observations (e.g. \citealt{winters00, gautschy04, nowotny10, nowotny11, sacuto11}), some fundamental open questions remain for the case of M-type AGB stars. As shown by \citet{woitke06}, silicate grains have to be basically iron-free to form and survive in the close vicinity of the star, resulting in insufficient absorption cross sections in the near-IR to drive an outflow. As a response to these findings, \citet{hoefner08} proposed an alternative solution, involving thermodynamic conditions where iron-free silicate grains are able to reach sizes in the range of 0.1-1.0\,$\mu$m in the close circumstellar environment (2 to 3 stellar radii). State-of-the-art radiation-hydrodynamic models demonstrate that forsterite particles (Mg$_{2}$SiO$_{4}$) in this size range have high enough radiative scattering cross-sections to drive an outflow \citep{hoefner08} and that the resulting visual and near-IR colors agree with observations \citep{bladh11, bladh12b}. Recently, aperture-masking polarimetric interferometry with the NAOS-CONICA VLT focal instrument at near-IR wavelengths has revealed the presence of scattered flux resulting from 0.3\,$\mu$m sized silicate grains in the close environment (about 2 stellar radii) of three low mass-loss, M-type AGB stars \citep{norris12}, giving strong observational support to the theoretical prediction by \citet{hoefner08}.\\

Because of the poor angular resolution of the IRAS, ISO, and SPITZER satellites (larger than 2$\arcsec$), the mid-infrared excess of AGB stars observed with these instruments is dominated by the thermal emission from the extended dusty circumstellar regions (above 100 stellar radii) where iron-rich silicate dust particles (Mg-Fe-silicates in the form of olivine and pyroxene) are thermally stable (e.g. \citealt{woitke06}). A clear way to disentangle iron-free and iron-rich silicate dust emission is to spatially resolve the innermost parts of the stellar envelope. High angular resolution observations in the mid-infrared are very attractive for this purpose, as they probe the close circumstellar region inside five stellar radii (e.g. \citealt{ohnaka04}), where only iron-free silicate particles can form and resist evaporation due to radiative heating \citep{woitke06}. By modeling the 10\,$\mu$m interferometric emission with self-consistent dynamic atmosphere and wind models, we test the theoretical predictions of \citet{hoefner08} and give additional observational constraints regarding winds driven by photon scattering on iron-free silicate grains.

\object{RT\,Vir} is an M-type (Sp.Type M8III), semi-regular variable star and one of the brightest water maser sources (e.g. \citealt{imai03, richards11}). The mass-loss rate of the star, found from water maser emission and CO line emission, is around 1 to 5$\times$10$^{-7}$ M$_{\odot}$\,yr$^{-1}$ \citep{bowers94,olofsson02,imai03}. Measurement values of the gas velocity vary from 6.2\,km\,s$^{-1}$ \citep{gonzalez03} in the SiO emitting region ($\sim$\,400 AU away from the star) to 7.8\,km\,s$^{-1}$ \citep{olofsson02} in the outer CO emitting region. Gas velocities up to 11\,km\,s$^{-1}$ \citep{richards11} have been measured in the water maser emitting region ($\sim$\,5 to 25 AU away from the star). 
The mid-infrared excess observed between 9\,$\mu$m and 13\,$\mu$m in the ISO/SWS data (see Fig.\,\ref{Flux-MIDI-ISO}) indicates that the star is surrounded by a dust shell. The pulsation period of the star has been estimated to 155 days by \citet{kholopov85} and reevaluated to 375 days by \citet{imai97} using the AAVSO visual lightcurve with magnitudes varying from V=7.7 to 9.7. The Hipparcos distance to the star is 135$\pm$15\,pc. The period-luminosity relation of \citet{whitelock08} gives a distance of 226\,pc, in very good agreement with the distance of 220\,pc derived by \citet{imai03} using both the statistical parallax, and model-fitting methods for the maser kinematics. We assume in the following that the distance of 220\,pc determined by \citet{imai03} is the most reliable, since the Hipparcos distance could suffer from a systematic bias due to photocenter displacement of the star \citep{chiavassa11}. The effective temperature of the star is estimated to 2900\,K (see Sect.\,\ref{MARCS}), while the stellar diameter has been evaluated to 12.4\,mas in the K and K' bands \citep{monnier04,mennesson02} and to 16.2\,mas in the L' band \citep{chagnon02,mennesson02}. Assuming that the uniform disk (UD) diameter in the K band is a good approximation of the continuum photospheric diameter \citep{scholz87}, and that the distance is 220\,pc and the effective temperature is 2900\,K, the luminosity of \object{RT\,Vir} is about 5500\,L$_{\odot}$.\\ 
 
The MIDI mid-infrared spectro-interferometric observations of \object{RT\,Vir} are presented in Sect.\,\ref{obs}. Section \ref{morphology} gives a first morphological interpretation of the object, based on the N-band interferometric data. The determination of the most appropriate dynamic model atmosphere for \object{RT\,Vir} is described in Sect.\,\ref{dyn}. In Sect.\,\ref{dust}, we aim at constraining the dust chemistry in the inner wind region of \object{RT\,Vir}, in order to test the theoretical predictions of \citet{hoefner08}. In Sect.\,\ref{discr} an alternative geometry is explored. Finally, the conclusions and future perspectives are given in Sect.\,\ref{concl}. 

\section{Spectro-interferometric observations}
\label{obs}

The Very Large Telescope Interferometer (VLTI) of ESO's Paranal Observatory was used with MIDI, the mid-infrared ($\lambda$=8-13 $\mu$m) interferometric recombiner \citep{leinert03}. MIDI combines the light of two telescopes and provides single-dish acquisition images, calibrated spectra, visibilities, and differential phases in the N-band atmospheric window. All the observations, with the exception of observation \#7 (see Table~\ref{journal-MIDI-AT}) made in \textit{HIGH-SENS} mode, were performed in the \textit{SCI-PHOT} mode of MIDI, meaning that photometry and interferometry were recorded simultaneously. The GRISM mode was used, providing a spectral resolution of about 230.

The observations of \object{RT\,Vir} were conducted in 2009 and 2011 with the VLT auxiliary telescopes (ATs) G0-H0, D0-H0, A0-G1, and A0-K0. These configurations provide projected baselines ranging from 128\,m down to 30\,m, probing spatial scales slightly above the stellar photosphere ($\sim$1.5 stellar radii) to regions located above the dust condensation zone ($\sim$6.5 stellar radii).

\begin{table*}[tbp]
\caption{\label{journal-MIDI-AT}Journal of the MIDI Auxiliary Telescopes observations of \object{RT\,Vir} with its main calibrator.} 
\begin{minipage}[h]{20cm}
\centering
\begin{tabular}{cccccccccc}\hline\hline
{\tiny \#} & {\tiny Star} & {\tiny UT date \& Time} & {\tiny $\phi_{\rm V}$ (Cycle)} & {\tiny Config.} & {\tiny Baseline length [m]} & {\tiny PA [deg]} & {\tiny Seeing [arcsec]} \\ \hline
{\tiny 1} & {\tiny \object{RT\,Vir}} & {\tiny 2009-03-18 05:03:28} & {\tiny 0.15 (1)} & {\tiny D0-H0} & {\tiny 59} & {\tiny 74} & {\tiny 1.7} \\ 
{\tiny Cal} & {\tiny \object{HD120323}} & {\tiny 2009-03-18 05:23:40} & {\tiny \ldots} & {\tiny -} & {\tiny \ldots} & {\tiny \ldots} & {\tiny 1.5} \\
{\tiny 2} & {\tiny \object{RT\,Vir}} & {\tiny 2009-03-18 06:03:49} & {\tiny 0.15 (1)} & {\tiny D0-H0} & {\tiny 63} & {\tiny 73} & {\tiny 1.0} \\ 
{\tiny Cal} & {\tiny \object{HD120323}} & {\tiny 2009-03-18 06:26:04} & {\tiny \ldots} & {\tiny -} & {\tiny \ldots} & {\tiny \ldots} & {\tiny 0.8} \\
{\tiny 3} & {\tiny \object{RT\,Vir}} & {\tiny 2009-06-03 01:33:04} & {\tiny 0.35 (1)} & {\tiny G0-H0} & {\tiny 32} & {\tiny 73} & {\tiny 1.0} \\ 
{\tiny Cal} & {\tiny \object{HD120323}} & {\tiny 2009-06-03 02:00:57} & {\tiny \ldots} & {\tiny -} & {\tiny \ldots} & {\tiny \ldots} & {\tiny 1.1} \\
{\tiny 4$^{\dag}$} & {\tiny \object{RT\,Vir}} & {\tiny 2009-07-01 00:26:23} & {\tiny 0.43 (1)} & {\tiny A0-K0} & {\tiny 126} & {\tiny 72} & {\tiny 2.4} \\ 
{\tiny Cal} & {\tiny \object{HD120323}} & {\tiny 2009-07-01 00:03:48} & {\tiny \ldots} & {\tiny -} & {\tiny \ldots} & {\tiny \ldots} & {\tiny 2.0} \\ \hline
{\tiny 5} & {\tiny \object{RT\,Vir}} & {\tiny 2011-02-08 08:29:37} & {\tiny 0.00 (3)} & {\tiny A0-G1} & {\tiny 89} & {\tiny 113} & {\tiny 0.9} \\ 
{\tiny Cal1} & {\tiny \object{HD120323}} & {\tiny 2011-02-08 08:06:59} & {\tiny \ldots} & {\tiny -} & {\tiny \ldots} & {\tiny \ldots} & {\tiny 0.9} \\
{\tiny Cal2} & {\tiny \object{HD120323}} & {\tiny 2011-02-08 08:55:11} & {\tiny \ldots} & {\tiny -} & {\tiny \ldots} & {\tiny \ldots} & {\tiny 1.0} \\
{\tiny 6} & {\tiny \object{RT\,Vir}} & {\tiny 2011-03-06 06:10:26} & {\tiny 0.06 (3)} & {\tiny G0-H0} & {\tiny 30} & {\tiny 74} & {\tiny 1.2} \\ 
{\tiny Cal1} & {\tiny \object{HD120323}} & {\tiny 2011-03-06 05:42:11} & {\tiny \ldots} & {\tiny -} & {\tiny \ldots} & {\tiny \ldots} & {\tiny 1.2} \\
{\tiny Cal2} & {\tiny \object{HD120323}} & {\tiny 2011-03-06 06:35:04} & {\tiny \ldots} & {\tiny -} & {\tiny \ldots} & {\tiny \ldots} & {\tiny 1.2} \\
{\tiny 7} & {\tiny \object{RT\,Vir}} & {\tiny 2011-03-13 07:06:34} & {\tiny 0.08 (3)} & {\tiny A0-K0} & {\tiny 128} & {\tiny 73} & {\tiny 1.4} \\ 
{\tiny Cal1} & {\tiny \object{HD120323}} & {\tiny 2011-03-13 06:46:31} & {\tiny \ldots} & {\tiny -} & {\tiny \ldots} & {\tiny \ldots} & {\tiny 1.3} \\
{\tiny Cal2} & {\tiny \object{HD120323}} & {\tiny 2011-03-13 07:26:59} & {\tiny \ldots} & {\tiny -} & {\tiny \ldots} & {\tiny \ldots} & {\tiny 1.2} \\
\hline
\end{tabular}
\end{minipage}
Note 1: Calibrators used to calibrate the visibilities are given after the science target. The visual phase $\phi_{\rm V}$ of the corresponding observing cycle is calculated from Eq.~\ref{eq-lightcurve}. The configuration, the length and position angle of the projected baseline, and the seeing value are also indicated.\\
Note 2: $^{\dag}$ means that the data were not exploitable (see text).
\end{table*}

\subsection{MIDI visibilities and differential phases} 
\label{ic}

The data reduction software package MIA+EWS\footnote{\tt{http://www.mpia-hd.mpg.de/MIDISOFT/, http://www.strw.leidenuniv.nl/$\sim$nevec/MIDI/}} was used to calibrate the visibility data. We checked that the resulting calibrated visibilities differed less than 5\% to 10\% between the reductions performed with MIA and with EWS. Because EWS also provides differential phases, the result of EWS is presented in the remainder of the paper. Various instrumental and atmospheric effects can affect the fringe contrast of the science target. Calibrator stars with known characteristics and dimensions must be observed close in time and space from the science target to derive the response of the instrument, and properly calibrate the interferometric measurements (e.g. \citealt{cruzalebes10}). During the first period (P83), we obtained measurements of the calibrator \object{HD120323} (F$_{12\,\mu m}$=255.4\,Jy and $\oslash$=9.16$\pm$0.07 mas)\footnote{IRAS 12\,$\mu$m flux and angular diameter from the ESO/MIDI calibrator database, respectively} before or after the measurements of \object{RT\,Vir}, while in the second observing period (P86), measurements of the same calibrator were obtained before and after \object{RT\,Vir} (see Table~\ref{journal-MIDI-AT}). The data quality has been checked using the \textit{MIA Graphical User Interface} (e.g. \citealt{sacuto11, klotz12a}). Among the MIDI measurements, data set \#4 shows very broad dispersed power spectral densities (PSD) and fringe histogram (FH), probably caused by the bad atmospheric conditions during the night (seeing$>$2$\arcsec$). If the PSD and FH are very broad, a very significant fraction of the fringe power is distributed outside the integration range, leading to a systematic underestimation of visibility (e.g. \citealt{sacuto11}). The corresponding visibility measurements were therefore discarded. 

\begin{figure}[tbp]
\begin{center}
\includegraphics[width=9.0cm]{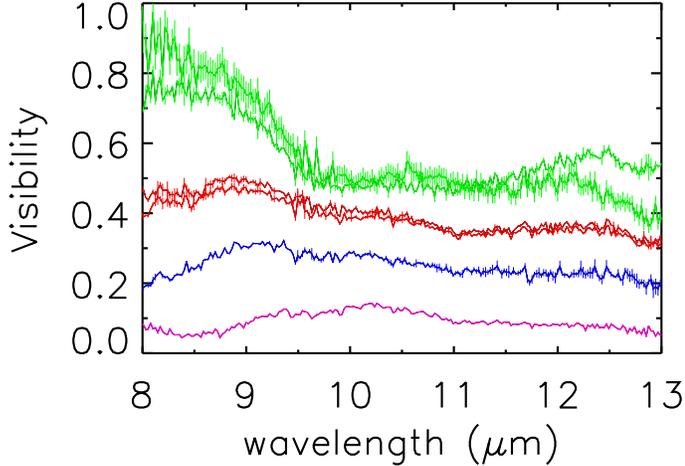}
\end{center}
\caption{Spectrally-dispersed MIDI calibrated visibilities of \object{RT\,Vir} observed over 5 nights between March 2009 and March 2011 (see Table\,\ref{journal-MIDI-AT}) at baselines 30(green lines), 60(red lines), 90(blue line), and 130\,m (purple line) from top to bottom, respectively.}
\label{VIS-MIDI}
\end{figure}

For P83, the calibrated visibilities from 2009 March 18 (data sets \#1 and \#2) were averaged over both bracketing calibrator observations obtained during that night, while error bars were derived from the standard deviation of the visibilities. This was not possible for the 2009 June 03 data (data set \#3), since no other calibrators were observed in the same mode (GRISM+SCI-PHOT) during that night, and a standard multiplicative error of 10\% was arbitrarily assumed for the visibilities \citep{chesneau07}. The calibrated visibilities from P86 were averaged over both bracketing calibrator observations obtained during each night, while error bars were derived from the standard deviation of the visibilities. Figure~\ref{VIS-MIDI} shows the MIDI-calibrated visibilities of \object{RT\,Vir} observed over the five nights (see Table~\ref{journal-MIDI-AT}).

\begin{figure}[tbp]
\begin{center}
\includegraphics[width=9.0cm]{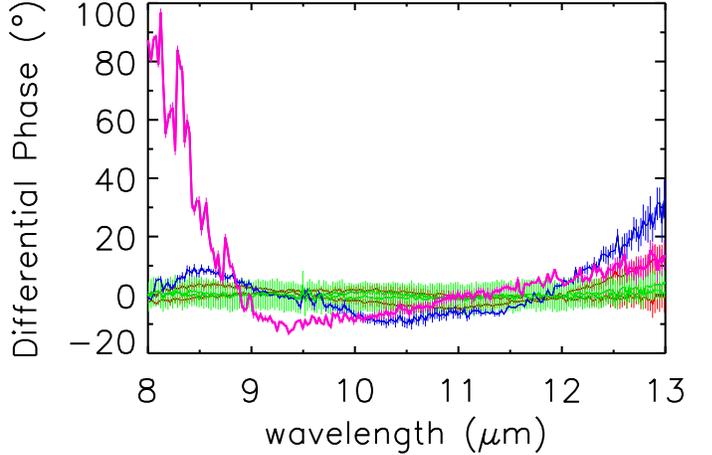}
\end{center}
\caption{Spectrally-dispersed MIDI calibrated differential phases of \object{RT\,Vir} (similar color-code as in Fig.\,\ref{VIS-MIDI}) observed over 5 nights between March 2009 and March 2011 (see Tab\,\ref{journal-MIDI-AT}). The bold purple line represents the largest 128\,m baseline measurements (data set \#7).}
\label{PHASE-MIDI}
\end{figure}

Error bars for the calibrated differential phases were derived in the same way as for the visibilities. 
Because the humidity and seeing levels were stable during 2009 June 03, a typical error of $\pm$5$^{\circ}$ was used for the data set \#3 differential phases \citep{ohnaka08}. Figure~\ref{PHASE-MIDI} shows the MIDI calibrated differential phases of \object{RT\,Vir} observed over the five nights (see Table~\ref{journal-MIDI-AT}). It is clear from this figure that the 128\,m baseline differential phases show a significant deviation from zero in the 8-9\,$\mu$m band\footnote{Differential phase measurements for wavelengths longer than 12.0\,$\mu$m are more uncertain owing to the increasing effect of the atmospheric water vapor and edge band noise. Therefore, the differential phases larger than the upper-limit uncertainty seen between 12.0 and 13.0\,$\mu$m for data set \#5 are probably not real.}, greater than the expected upper-limit uncertainty of 10$^{\circ}$. A similar differential phase signature was found by \citet{paladini12} in the close environment of the C-type star R\,For.

\subsection{MIDI and ISO/SWS spectra} 
\label{fc}

MIDI spectra are derived from the average of the calibrated spectra close in visual phase within a given observing cycle, while the error bars are determined from the standard deviation of the calibrated spectra. Since the calibrated spectrum of data set \#7 leads to an unusual spectral signature that is different from the other calibrated spectra, we decided to discard this measurement together with data set \#4 (see Sect.~\ref{ic}).\\ 

In the following, the visual phase of the star is estimated from its AAVSO lightcurve \citep{henden09} according to

\begin{equation}
\label{eq-lightcurve}
\phi_{\rm V}=\frac{(t - T_{0}) \, \rm mod \, P}{\rm{P}},
\end{equation}
where $t$ is the observing time in Julian date, $T_{0}$=2\,454\,854 is the Julian date of the selected phase-zero point corresponding to the maximum light of the star ($\phi^{0}_{\rm V}$=0.0), and P=375 days is the pulsation period of the star \citep{imai97}. \\ 

\begin{figure}[tbp]
\begin{center}
\includegraphics[width=9.0cm]{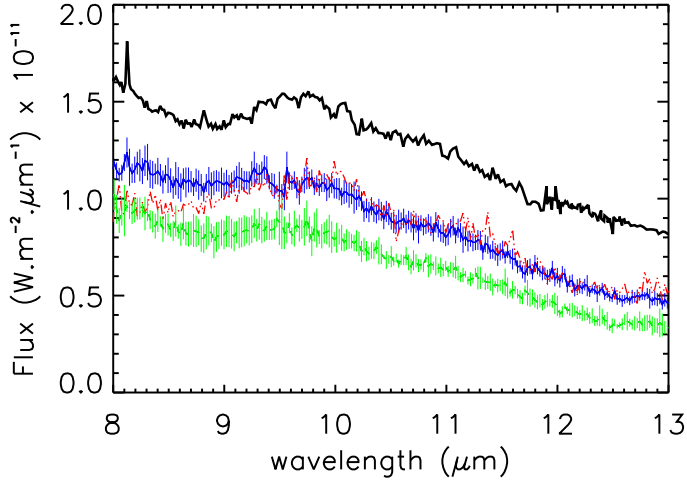}
\end{center}
\caption{Comparison of the ISO/SWS spectrum of \object{RT\,Vir} (black thick line) and the MIDI calibrated spectra of \object{RT\,Vir} at phase $\phi_{\rm V}$=0.15 (Cycle 1) (blue solid line + error bars), phase $\phi_{\rm V}$=0.35 (Cycle 1) (red dashed-dotted line), and at phase $\phi_{\rm V}$$\sim$0.0 (Cycle 3) (green dashed line + error bars).}  
\label{Flux-MIDI-ISO}
\end{figure}

Figure~\ref{Flux-MIDI-ISO} compares the calibrated MIDI spectra for both periods (2009/2011) with the ISO/SWS spectrum of \object{RT\,Vir} observed 1996 July 20. The ISO/SWS absolute flux level of \object{RT\,Vir} was scaled to the IRAS 12 and 25\,$\mu$m photometry of the star. The MIDI flux level is $\sim$50\% lower than that of ISO/SWS. Two possibilities could account for this difference. The first one is that the beam of an AT ($\sim$1.1$\arcsec$ at 10~$\mu$m) is small compared to that of the IRAS satellite\footnote{Keeping in mind that both 12 and 25\,$\mu$m IRAS photometry were used to scale the ISO spectrum.}. Because of the relatively short distance to \object{RT\,Vir} (220 pc), and the large extent of its dust envelope, some of the N-band flux might be lost outside the AT field-of-view. A second possibility that cannot be excluded is that the flux level has changed since the IRAS observation (performed in 1983), owing to cycle-to-cycle stellar variability. The AAVSO lightcurve shows an increase in the star visual amplitude by a factor of about 4 from the IRAS to the MIDI observations. Moreover, there is a $\sim$20\% decrease in the MIDI flux from P83 (cycle 1) to P86 (cycle 3) (see Fig.\ref{Flux-MIDI-ISO}). Since the P83 and P86 periods are close in phase, this loss in brightness can be due to a cycle-to-cycle variability of the object.

\section{Morphology of \object{RT\,Vir} in the mid-infrared}
\label{morphology}

To constrain the morphology of the object, we used the software GEM-FIND (GEometrical Model Fitting for INterferometric Data; \citealt{klotz12b}). For the best reliability, the geometrical fitting requires the largest possible \textit{uv} coverage \citep{klotz12b}. To check for interferometric variability of the source (which may compromise the use of the full data set for the fit), we can compare the visibility measurements of data sets \#3 and \#6. The two data sets have very similar baseline lengths and position angles, but were observed during two different cycles (see Table~\ref{journal-MIDI-AT}). The comparison between the two green upper curves of Fig.~\ref{VIS-MIDI} shows that no significant interferometric variability is found\footnote{With the exception of wavelengths longward of 12.3\,$\mu$m that can be due to an underestimation of the uncertainties for data sets \#3 and/or \#6 because of edge band noise within this wavelength range or because of the small baseline length difference (2\,m) between both data sets.}, unlike what is found for the MIDI N-band flux measurements (see Fig.~\ref{Flux-MIDI-ISO}), which are dominated by the emission from the extended circumstellar regions. The fact that no significant cycle-to-cycle variability is found for the 30\,m MIDI visibilities, which probe the inner atmospheric regions, reveals different motions of the circumstellar environment from one spatial scale to the other. 

To account for the central star, we used a spherical UD model. We used a spherical Gaussian distribution to reproduce the optically thin dusty envelope of the star. To account for the deviation from zero-differential phase in the 8-9\,$\mu$m wavelength region of the 128\,m baseline measurement (see Fig.~\ref{PHASE-MIDI}), we added an offset Dirac delta distribution representing an unresolved component close to the star. The diameter of the UD representing the central star and the FWHM of the Gaussian distribution representing the dusty envelope are both wavelength dependent, in order to account for the opacity variation in the close stellar atmosphere and of the dust over the N-band range. Both the flux ratio of the envelope to the central star and the flux ratio of the unresolved component to the central star are also wavelength dependent, in order to account for the flux variation across the N-band for all three components.

\begin{figure}[tbp]
\begin{center}
\includegraphics[width=9.0cm]{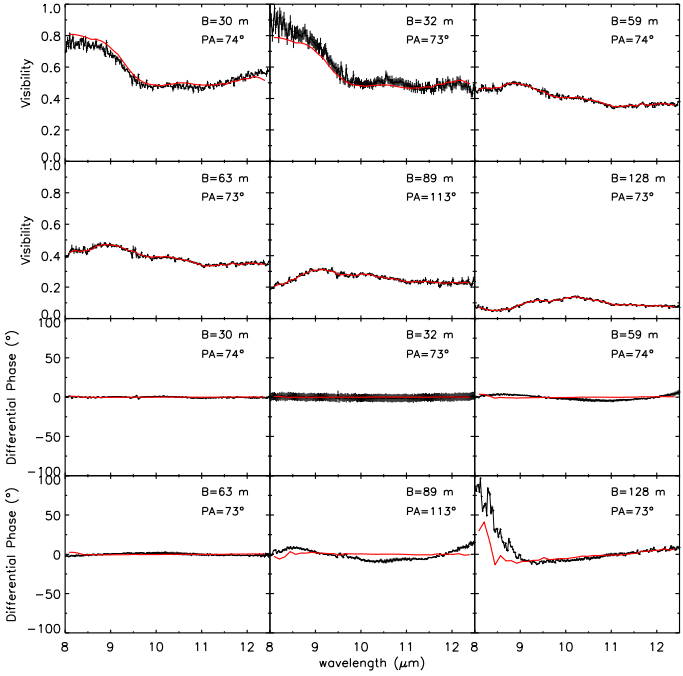}
\end{center}
\caption{Best fit of the UD+Gauss+Dirac geometrical model (red solid lines) on the calibrated visibilities and differential phases of \object{RT\,Vir} (black solid lines).}
\label{geo}
\end{figure}

\begin{figure}[tbp]
\begin{center}
\includegraphics[width=9.0cm]{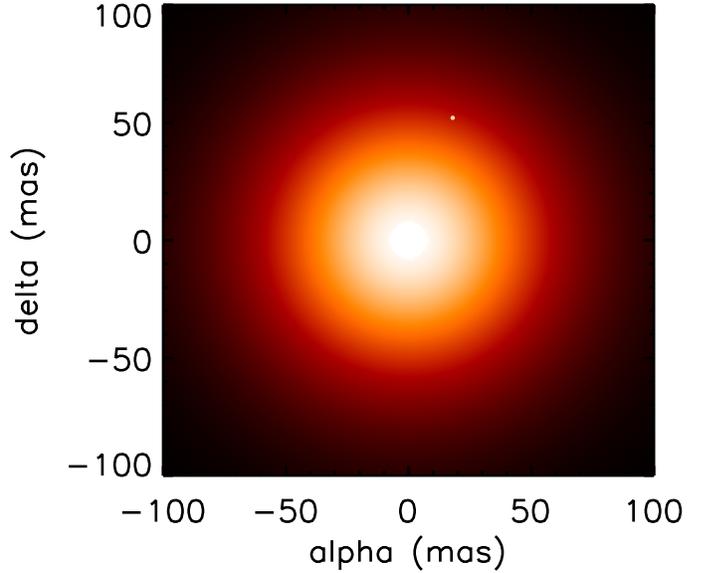}
\end{center}
\caption{Normalized intensity distribution of the best-fitting geometrical model at 10.5\,$\mu$m. The offset white circle represents the unresolved component. North is up and east is left.}
\label{Int_RT_Vir}
\end{figure}

\begin{figure}[tbp]
\begin{center}
\includegraphics[width=9.0cm]{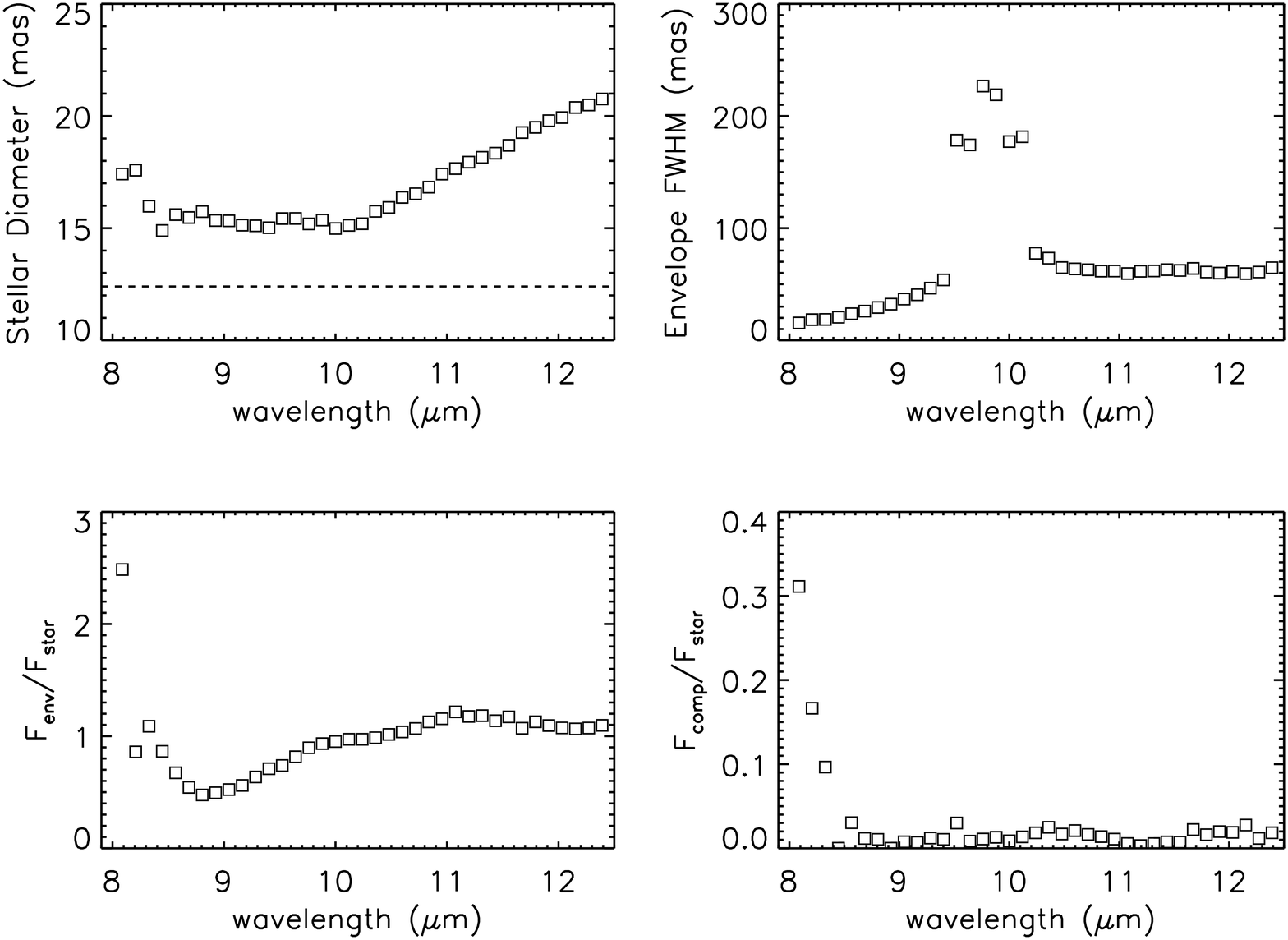}
\end{center}
\caption{Wavelength-dependent parameter values of the best-fitting geometrical model. The upper left panel shows the stellar diameter. The K-band diameter of the star is superimposed as a dashed line. The upper right panel represents the FWHM of the dusty envelope. The lower left panel represents the flux ratio of the dusty envelope over the central star, while the lower right panel shows the flux ratio of the unresolved component over the central star.}
\label{param_geo}
\end{figure}

The fitting strategy is the same as described in \citet{klotz12b}. Figure\,\ref{geo} presents the best-fitting geometrical model together with the calibrated MIDI visibility and differential phase measurements of \object{RT\,Vir}. Figure\,\ref{Int_RT_Vir} shows the normalized intensity distribution of the best model at 10.5\,$\mu$m. Figure\,\ref{param_geo} gives the wavelength-dependent parameters of the best-fitting model.\\ 

All visibility and 30 to 89\,m baseline differential phase measurements are very well reproduced by the model. The similar trend of the model and the 128\,m baseline differential phases in the 8-9\,$\mu$m range provides one simple solution to the nature of the asymmetry\footnote{Keeping in mind that due to the lack of spatial coverage, more complex morphologies cannot be excluded.}. It is assumed here that the asymmetry corresponds to an unresolved component having a flux of about 13\,Jy from 8 to 8.5\,$\mu$m (and about 2 Jy through the rest of the N-band), located at a projected distance of about nine stellar radii (12\,AU away from the central star). However, the position of the unresolved component is not well constrained owing to the very limited uv-coverage. The result from the geometrical modeling shows that the unresolved component does not disrupt the sphericity of the close circumstellar environment (see Fig.\,\ref{Int_RT_Vir}), and could therefore be related to, e.g., an H$_{2}$O or SiO maser clump (i.e. \citealt{richards11,richards12}).

The overall larger stellar diameter (compared to the 12.4 mas K/K'-bands diameter) in the N-band (see in the upper left panel of Fig.\,\ref{param_geo}) is probably caused by the presence of warm molecular layers of H$_{2}$O and SiO (e.g. \citealt{ohnaka05}). Furthermore, the increase in the stellar diameter from 16\,mas (1.3 stellar radii) to 21\,mas (1.7 stellar radii) between 10.3 and 12.5\,$\mu$m is possibly correlated to the presence of Al$_{2}$O$_{3}$ (corundum), which is optically efficient around 11\,$\mu$m (see Fig.\,\ref{qeff_mir}), and has already been suggested to explain similar features seen in mid-infrared interferometric measurements of other comparable sources (e.g. \citealt{wittkowski07,zhao12}).  

The FWHM of the Gaussian distribution representing the dusty envelope increases from 16\,mas (1.3 stellar radii) to 54\,mas (4.3 stellar radii) between 8 and $\sim$9.4\,$\mu$m. This increase is caused by a silicate dust opacity effect. The modeling also shows a large increase in the FWHM between 9.4 and 10.4\,$\mu$m with a peak reaching 227\,mas (18 stellar radii) around 9.8\,$\mu$m. Such an increase is very likely related to an extended dusty circumstellar envelope composed of silicate grains. An extra shell of silicate dust also needs to be included in the dynamic modeling to explain the 30 and 32\,m baseline visibility measurements (see Sect.\,\ref{discr}). Finally, the FWHM shows a constant size of about 60 mas (5 stellar radii) between 10.4 and 12.5\,$\mu$m.\\ 

This first morphological interpretation already gives us some clues to the presence of aluminum and silicate dust in the close environment (1.3 to 4.3 stellar radii) of \object{RT\,Vir}. Detailed theoretical modeling \citep{woitke06} suggests that silicate grains forming on such spatial scales have to be iron-free not to evaporate from heating by the radiation field. 

The MIDI differential phases show the presence of an asymmetry in the stellar vicinity. However, since the asymmetry does not affect the overall sphericity of the close circumstellar environment, this allows the use of the spherically symmetric dynamical model to investigate the presence of aluminum and iron-free silicate dust.

\section{The dynamic model atmosphere of \object{RT\,Vir}}
\label{dyn}

The dynamic model atmospheres (called DMAs hereafter) used are described in detail in \citet{hoefner08}. They contain a consistent time-dependent description of grain growth for pure forsterite grains, and are based on radiative transfer in 319 wavelength points, taking size-dependent dust grain opacity into account. Each dynamic model starts from a hydrostatic initial structure, and the effects of stellar pulsation are simulated by a variable inner boundary just below the stellar photosphere (piston model). The dynamic code provides the temporally varying temperature and density distribution of the object. Based on the radial structure at given instances in time, atomic and molecular opacities are computed using the COMA code\footnote{This version of the COMA code does not take dust scattering into account.} \citep{aringer09}. COMA solves the ionization and chemical equilibrium equations at a given temperature and density combination for a set of atomic abundances assuming LTE. The calculated opacities are then introduced into a spherically symmetric radiative transfer code to calculate the emergent intensity distribution at any desired spectral resolution. This intensity distribution is used to derive the synthetic spectrum and visibility profile, which are compared to the observational data of the star.

\subsection{Strategy}

The strategy of selecting the most appropriate DMA for \object{RT\,Vir} is similar to the one used by \citet{sacuto11} for \object{R\,Scl}. 

\begin{enumerate}

\item Constrain the best set of stellar parameters, i.e. surface gravity, mass, effective temperature, and metallicity, which are used for the initial hydrostatic structure of the DMA. This is done by fitting hydrostatic MARCS models to the broadband photometric and ISO/SWS spectrometric measurements of the star. \\ 

\item Constrain the additional input parameters of the DMA, i.e., the amplitude of the piston and abundance of seed particles. This is done in two steps:\\

a. Compute a sample of DMAs using the best-fit stellar parameters derived from the hydrostatic modeling. Different velocity amplitudes of the piston and abundances of seed particles are used to produce models with different mass-loss rates and wind velocities. Models producing mass-loss rates and wind velocities in agreement with the values found for \object{RT\,Vir} in the literature are selected for further analysis. \\ 

b. Among the set of models selected in the previous step, constrain the best velocity amplitude of the piston and abundance of seed particles, by comparing the time-dependent synthetic dust-free spectra to the broadband photometric and ISO/SWS spectrometric measurements.

\end{enumerate}

These points are detailed in Sects.\,\ref{MARCS}, \ref{DMA_creation}, and \ref{dma_constraints}, respectively.

\subsection{Determination of the stellar parameters of \object{RT\,Vir}}
\label{MARCS}

The stellar parameters are determined by comparing hydrostatic stellar atmosphere models to the spectrophotometric data. The hydrostatic model that provided the best fit to the broadband B, V (simbad database) and J, H, and K photometric data from \citet{Kerschbaum94} and the ISO/SWS spectra of the star up to 5\,$\mu$m was chosen. This limit of 5\,$\mu$m is fixed because the contributions from atmospheric molecular and dust emission become non-negligible at longer wavelengths for optically thin, \emph{warm, dusty objects} like \object{RT\,Vir} \citep{kraemer02}, and that cannot be accounted for by the hydrostatic modeling.

\begin{figure}[tbp]
\begin{center}
\includegraphics[width=9.0cm]{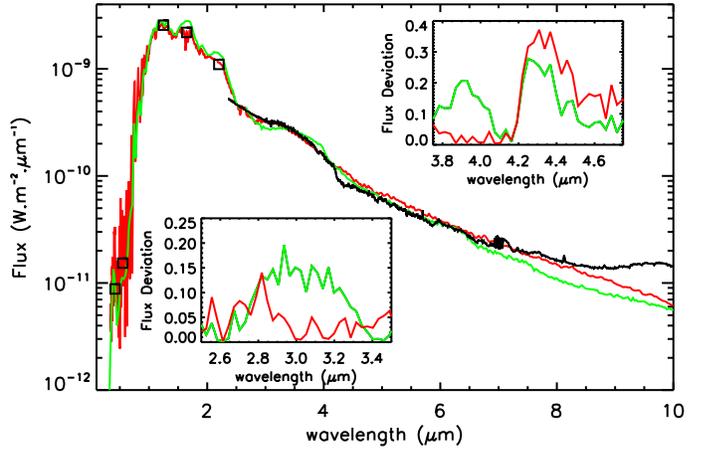}
\caption{Best-fitting hydrostatic MARCS atmospheric spectrum (green solid line) and best-fitting dust-free spectrum of the DMA~1 (red solid line) at bolometric phase $\phi_{\rm bol}$=0.59, together with the broadband photometric data (black squares) and ISO/SWS spectrum (black solid line) of \object{RT\,Vir}. The insets present the flux deviation of the hydrostatic MARCS spectrum (green solid line) and the flux deviation of the dust-free spectrum of the DMA~1 (red solid line) from the ISO/SWS spectrum around 3\,$\mu$m and 4.25\,$\mu$m, respectively.}
\label{MARCS_HYDRODYNnodustonephase_phot_ISO}
\end{center}
\end{figure}

A least-squares minimization fitting was done using a grid of MARCS models \citep{gustafsson08}, which has a basic sun-like chemical composition as listed by \citet{grevesse07}. The effective temperature of the star was varied from 2800 to 3200~K ($\Delta$T$_{\rm eff}$=100 K) for metallicity values of -0.25, 0.0, and 0.25. The surface gravity log\,\textit{g}=-0.5 of the model was fixed according to a stellar mass M=1\,M$_{\odot}$, the distance d=220\,pc deduced by \citet{imai03}, and the angular diameter of the central star of 12.4\,mas derived from K-band long-baseline interferometry by \citet{monnier04}. Table~\ref{stellar-parameters} summarizes the stellar atmosphere characteristics of \object{RT\,Vir} that was adopted, as well as deduced, from the best-fitting hydrostatic model. Figure~\ref{MARCS_HYDRODYNnodustonephase_phot_ISO} presents the best-fitting MARCS model, together with the spectrophotometric data. A discrepancy between the model and the measurements is seen around 3\,$\mu$m corresponding to the H$_{2}$O and OH absorption bands, as well as around 4.25\,$\mu$m corresponding to the CO$_{2}$ absorption band. The photospheric part of these bands provided by the hydrostatic model may not be enough to explain the observed features that could originate in the extended atmospheric environment. Finally, the temperature of T$_{\rm eff}$=2900\,K leads to a luminosity of L=5500\,L$_{\odot}$ for the star.

\begin{table}[h]
\caption{Stellar atmospheric parameters of \object{RT\,Vir} fixed and deduced from fitting a hydrostatic model to the spectrophotometric measurements.}
\label{stellar-parameters}
\centering
\begin{tabular}
[c]{ccc} \hline\hline
Method & Parameter & Value \\ \hline
& Stellar diameter: $\theta$ (mas) & 12.4 \\ 
Fixed & Stellar mass: $M$ (M$_{\odot}$) & 1 \\
& Surface gravity: log\,$g$ & -0.5 \\\hline
& Effective temperature: T$_{\rm eff}$ (K) & 2900 \\ 
Deduced from the fit & Metallicity: Z (Z$_{\odot}$) & 1 \\\hline
\end{tabular}
\end{table}

Now that the stellar parameters of \object{RT\,Vir} have been determined, they are used for the initial hydrostatic structure to compute the corresponding DMAs having mass-loss rates and wind velocities in agreement with the ones found in the literature.

\subsection{Constraining the additional parameters for the DMA}
\label{dyn_param}

\subsubsection{Dynamic model atmosphere sample}
\label{DMA_creation}

In addition to the stellar parameters determined from the best fit of the hydrostatic model to the spectrophotometric measurements (see Sect.~\ref{MARCS}), the dynamical computation requires three other free parameters to be determined: (i) the abundance of seed particles per H atoms n$_{\rm gr}$/n$_{\rm H}$, (ii) the velocity amplitude of the piston $\Delta$u$_{\rm p}$, and (iii) the period of the piston P$_{\rm mod}$. \\

While the period of the piston is fixed to P$_{\rm mod}$=375\,days, i.e., the pulsation period of the star \citep{imai97}, different values of n$_{\rm gr}$/n$_{\rm H}$ and $\Delta$u$_{\rm p}$ are used, resulting in the development of a wind, characterized by a mass-loss rate $\dot{M}$ and a wind velocity $u$. The range of values for $\Delta$u$_{\rm p}$ and n$_{\rm gr}$/n$_{\rm H}$ are constrained by the values of $\dot{M}$ and $u$ derived from other studies: 1 to 5$\times$10$^{-7}$ M$_{\odot}$\,yr$^{-1}$ \citep{bowers94,olofsson02,imai03} and 6.2 to 11\,km\,s$^{-1}$ \citep{olofsson02,gonzalez03,richards11}.\\

The parameters and resulting wind properties of the detailed DMAs are listed in Table~\ref{Dyn_Mod}. 

\begin{table}[h]
\caption{DMAs parameters and resulting wind properties.} 
\label{Dyn_Mod}
\begin{minipage}[h]{10cm}
\begin{tabular}
[c]{ccc|cc} \hline\hline
DMA \# & n$_{\rm gr}$/n$_{\rm H}$ & $\Delta$u$_{\rm p}$ & $<$\.{M}$>$  & $<$u$>$ \\ 
 &  & [km\,s$^{-1}$] & [M$_{\odot}$\,yr$^{-1}$] & [km\,s$^{-1}$] \\ \hline
1 & 3$\times$10$^{-15}$ & 3 & 1.9$\times$10$^{-7}$ & 6.6 \\ 
2 & 3$\times$10$^{-15}$ & 3.5 & 6.7$\times$10$^{-7}$ & 11.4 \\ 
3 & 1$\times$10$^{-15}$ & 4 & 7.7$\times$10$^{-7}$ & 7.1 \\ 
4 & 1$\times$10$^{-15}$ & 5 & 1.4$\times$10$^{-6}$ & 11.5 \\ \hline
\end{tabular}
\end{minipage}
The fundamental stellar parameters of the DMAs presented here are the ones listed in Table~\ref{stellar-parameters}.\\
\footnotemark[1]{n$_{\rm gr}$/n$_{\rm H}$ and $\Delta$u$_{\rm p}$ are inputs values of the model, while $<$\.{M}$>$ and $<$u$>$ are output values.}
\end{table}

\subsubsection{Constraining the DMA of \object{RT\,Vir}}
\label{dma_constraints}

To select the most appropriate DMA for \object{RT\,Vir}, among the ones given in Table~\ref{Dyn_Mod}, we compare the time-dependent spectra computed without dust opacities of each DMA~1/2/3/4 to the spectrophotometric data below 8\,$\mu$m. This limit is fixed because of the ability of the dynamical models, in contrast to the hydrostatic model, to self-consistently develop extended atmospheric molecular structure of H$_{2}$O, SO$_{2}$, and SiO typically emitting between 5 and 8~$\mu$m. Beyond 8~$\mu$m, the dust emission is non-negligible, so addition of dust opacities in the radiative transfer calculation is required (see Sect.\,\ref{dust}).

We investigated 46 snapshots from each model presented in Table~\ref{Dyn_Mod}, equidistant in phase\footnote{The phase of the DMA corresponds to the bolometric phase $\phi_{\rm bol}$ derived from its bolometric lightcurve \citep{nowotny05b}.} and distributed over three consecutive pulsation cycles. The additional parameters, n$_{\rm gr}$/n$_{\rm H}$ and $\Delta$u$_{\rm p}$, were derived by fitting the average dust-free synthetic spectrum for each DMA (averaged over all phases and cycles) to the spectrophotometric data below 8\,$\mu$m. Another test was done by finding the individual dust-free spectrum that presents the best agreement with the spectrophotometric data for a given bolometric phase. 
In both cases, DMA~1 gives the best agreement with the spectrophotometric data (see Fig.\,\ref{MARCS_HYDRODYNnodustonephase_phot_ISO}). Spectra of DMA 2/3/4 having a higher $<$\.{M}$>$, overestimate the flux compared to the spectrophotometric measurements. While the best-fitting DMA~1 can reproduce the 3\,$\mu$m band, in contrast to the hydrostatic model, it is not able to reproduce the measurements around 4.25\,$\mu$m (CO$_{2}$ absorption band), as is also the case for the hydrostatic modeling (see Sect.\,\ref{MARCS}). Because the contribution of extended molecular layers is accounted for by the DMA, the reason for the 4.25\,$\mu$m discrepancy can be a deviation from chemical equilibrium in the low density outermost atmosphere (400-1000\,K), where the CO$_{2}$ feature is formed. In contrast to the hydrostatic MARCS models, Fig.\,\ref{MARCS_HYDRODYNnodustonephase_phot_ISO} shows that the best-fitting DMA~1 is able to reproduce the 6.5-8\,$\mu$m excess of H$_{2}$O, SO$_{2}$, and SiO emission from the extended atmosphere.

\section{Dust chemistry in the inner wind region of \object{RT\,Vir}} 
\label{dust}

A major point of this study is to test the theoretical predictions of \citet{hoefner08}, i.e., that winds of M-type AGB stars can be driven by photon scattering on iron-free silicate dust formed around two to three stellar radii. To do so, we compare the N-band spectro-interferometric measurements with the forsterite dust-rich spectrum and visibility profiles of the selected DMA~1 (see Sect.\,\ref{dma_constraints}).

\begin{figure}[tbp] 
\begin{center}
\includegraphics[width=8.5cm]{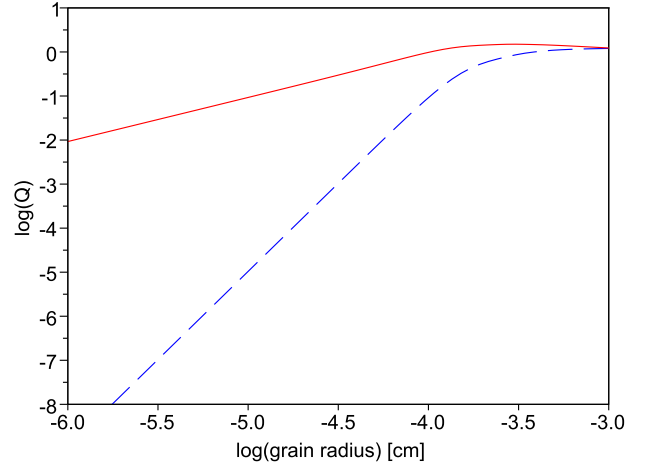} 
\caption{Absorption $Q_{abs}$ (red solid line) and scattering $Q_{sca}$ (blue dashed line) efficiencies of Mg$_{2}$SiO$_{4}$ as functions of grain radius at $\lambda$=9.8\,$\mu$m. Data for the refractive index are taken from \citet{jager03}. The efficiencies are calculated using the Mie theory for spherical particles (program BHMIE from \citet{bohren83} and modified by Draine, www.astro.princeton.edu/draine/scattering.html).}
\label{efficiency_10mic}
\end{center}
\end{figure}

The predicted mean grain size of DMA~1 is $<$$a_{\rm gr}$$>$=0.3\,$\mu$m. Since the scattering efficiency of forsterite grains with sizes below 1\,$\mu$m is lower than the absorption efficiency in the mid-infrared (see Fig.\,\ref{efficiency_10mic}), this allows the use of COMA (with no scattering included) for the comparison between the DMA and the MIDI interferometric data.

\subsection{Iron-free silicate grains as wind drivers} 

Using the temperature and density structures given by the dynamical calculation, the intensity distribution can be computed within the N-band wavelength range at the MIDI-GRISM spectral resolution (230) with the help of the radiative transfer code COMA for the 46 snapshots of DMA 1. Because the COMA code uses the same molecular and dust (forsterite) opacities, such as the ones included in the dynamical computation, the radiative transfer is done consistently. By Fourier transform of the intensity profiles for the different projected baselines of the observations (see Table\,\ref{journal-MIDI-AT}), we can derive the corresponding dispersed visibility profiles and compare them to the MIDI measurements of the star, following the same approach as described in \citet{sacuto11}. We then select the best-fitting bolometric phases of DMA~1 by least-squares minimization between the results of the model and the visibility measurements for periods 2009 (cycle 1) and 2011 (cycle 3). 

\begin{figure}[tbp]
\begin{center}
\includegraphics[width=9.0cm]{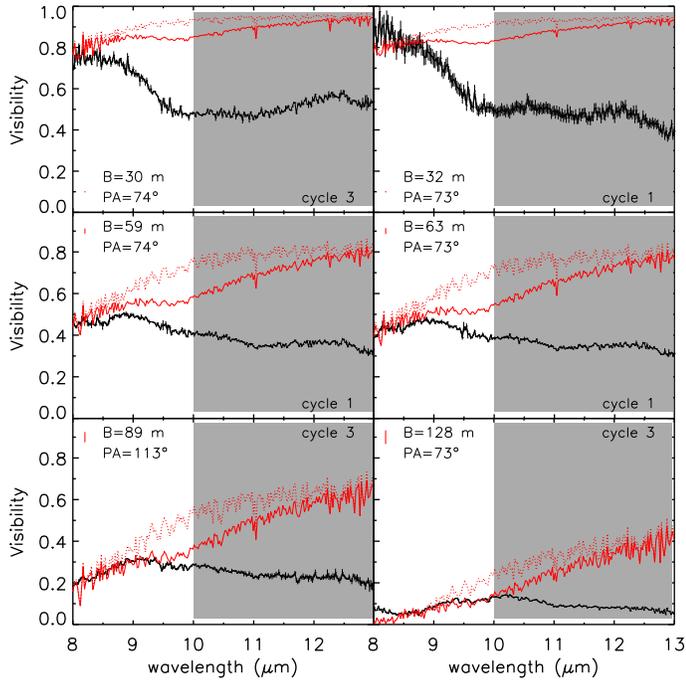}
\caption{Best-fitting synthetic visibility profiles with forsterite of DMA~1 at phase $\phi_{\rm bol}$=0.60 and $\phi_{\rm bol}$=0.86 (red solid lines), together with the MIDI measurements of the star (black solid lines) for periods 2009 (cycle 1) and 2011 (cycle 3), respectively. Red bars, next to the legend, represent the average phase dispersion of the model. Red dotted lines represent the corresponding dust-free visibility profiles at the same bolometric phases. Unshaded areas represent the silicate dust's main effective domain.}
\label{COMPA_VIS_DMApureMg2SiO4_MIDIdataRTVir}
\end{center}
\end{figure}

Figure\,\ref{COMPA_VIS_DMApureMg2SiO4_MIDIdataRTVir} shows the best-fitting synthetic visibility profiles with forsterite of DMA~1, together with the MIDI measurements. We show, for comparison, the visibility profiles computed without dust opacities for the same bolometric phases to demonstrate the impact of the silicate grains.\\

\begin{figure}[tbp]
\begin{center}
\includegraphics[width=9.0cm]{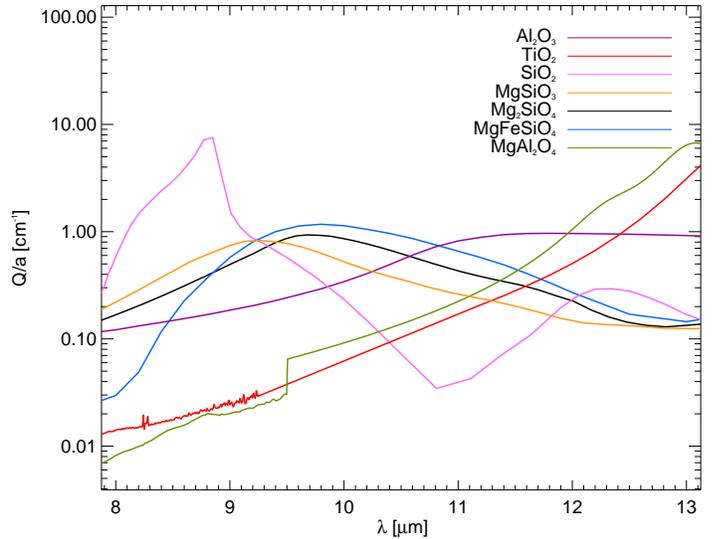}
\caption{Optical efficiency per grain radius Q$_{abs}$/a$_{gr}$ as a function of wavelength in the mid-infrared region for a selection of dust species likely to exist in the circumstellar environments of M-type AGB stars. Refractive index data were taken from \citet{begemann97} (Al$_2$O$_3$), \citet{zeidler11} (TiO$_2$, MgAl$_2$O$_4$), \citet{palik85} (SiO$_2$), \citet{jager03} (MgSiO$_3$ and Mg$_2$SiO$_4$), and \citet{dorschner95} (MgFeSiO$_4$).}
\label{qeff_mir}
\end{center}
\end{figure}

The unshaded areas of Fig.\,\ref{COMPA_VIS_DMApureMg2SiO4_MIDIdataRTVir} correspond to the wavelength region where silicate species show strong spectral features (see Fig.\,\ref{qeff_mir}). Figure\,\ref{COMPA_VIS_DMApureMg2SiO4_MIDIdataRTVir} reveals that the fully self-consistent DMA~1, producing forsterite dust particles of sizes $\sim$0.3\,$\mu$m, is in better agreement with the interferometric measurements than the dust-free model, especially for the 59 and 63\,m baselines, taking the average phase dispersion of the model into account shown by the bar next to the legend of each plot.

This result provides evidence for the presence of silicate dust in the region around two to three stellar radii probed by the 60\,m baseline (considering an angular diameter of the star of 12.4\,mas). This gives observational support, in addition to what is provided by \citet{norris12}, in favor of the theoretical predictions of \citet{hoefner08}, i.e., that winds can be driven by photon scattering on iron-free silicate grains formed in the close vicinity of M-type AGB stars. A further test for the presence of silicate dust around two to three stellar radii is discussed in Sect.\,\ref{add-confirm}.

\subsection{Exploring extra dust opacities}
\label{extra-dust}

In the dynamical computation, the formation of only one type of dust, i.e. pure forsterite (Mg$_{2}$SiO$_{4}$), is considered. However, it is likely that other types of dust can contribute to the enhancement of the opacity in the mid-infrared region. 
Figure\,\ref{COMPA_VIS_DMApureMg2SiO4_MIDIdataRTVir} reveals that the synthetic visibilities are above the measurements longward of 10\,$\mu$m, indicating that material strongly emitting around 11\,$\mu$m in the close stellar vicinity is missing in the radiative transfer computation. Figure\,\ref{qeff_mir} shows the efficiency per grain radius as a function of wavelength for a selection of oxygen-bearing dust species. From this plot, the only dust component that is optically efficient around 11\,$\mu$m is Al$_{2}$O$_{3}$ (corundum) dust. Therefore, to improve the modeling of the visibility measurements longward of 10\,$\mu$m, we try to compensate for the lack of opacity of the model around 11\,$\mu$m by adding alumina dust in the a-posteriori radiative transfer computation.\\

Additional dust species can be included a-posteriori in the COMA radiative transfer code, as was done by \citet{sacuto11} for SiC dust in the case of \object{R\,Scl}. The radiative transfer is done inconsistently in this case because for M-type AGB stars, the dynamical computation does not include other dust opacities than forsterite. However, because of the low abundance of Al, aluminum dust should not influence the process of wind driving in the dynamical computation \citep{bladh12a}. Nevertheless, aluminum dust, like corundum, is likely to contribute to the opacity enhancement around 11\,$\mu$m (see Fig.\,\ref{qeff_mir}) leading to better agreement between the model and the mid-infrared visibility measurements. 
Finally, interferometric measurements in the mid-infrared have already pointed to the presence of Al$_{2}$O$_{3}$ dust around two stellar radii in Mira and semi-regular M-type AGB stars (e.g. \citealt{wittkowski07,zhao12}). 
We therefore decided to add corundum in the a-posteriori COMA radiative transfer computation, using the small-particle limit approximation. 

Refractive indices of corundum were adopted from \citet{begemann97}, while the condensation temperature was fixed to $T_{\rm cond}^{\rm Al_{2}O_{3}}$=1400\,K (see e.g. \citealt{gail10}). Since aluminum dust is not included in the dynamical computation, we assumed that corundum follows the same temperature-density structure as the gas. The only free parameter left is the degree of condensation ($\eta_{\rm Al_{2}O_{3}}$) of Al condensing into corundum. We varied $\eta_{\rm Al_{2}O_{3}}$ from 10\% to 100\% (with $\Delta\eta_{\rm Al_{2}O_{3}}$=10\%) to minimize the difference between the model and the visibilities for all baselines. Even if the addition of Al$_{2}$O$_{3}$ dust increases the opacity and leads to a decrease in the visibility amplitude longward of 9\,$\mu$m, 100\% of Al condensing into corundum is not sufficient to reach the level of the visibility measurements. However, because the overall shape of the synthetic visibility profile is similar to the measurements longward of 9\,$\mu$m, we decided to increase the ratio above 100\% assuming the creation of more dust than available Al atomic material, i.e., exceeding the solar abundance of Al. Varying $\eta_{\rm Al_{2}O_{3}}$ from 200\% to 1000\% (with $\Delta\eta_{\rm Al_{2}O_{3}}$=100\%), we find the best agreement with the measurements for $\eta_{\rm Al_{2}O_{3}}$=500\%. 

From the measurements, a clear decrease in the visibility profile is revealed above 12.5\,$\mu$m. Going back to Fig.\,\ref{qeff_mir}, both MgAl$_{2}$O$_{4}$ (spinel) and TiO$_{2}$ (titanium dioxide) show an increase in their optical efficiency longward of 12.5\,$\mu$m. Since TiO$_{2}$ has a lower optical efficiency than MgAl$_{2}$O$_{4}$ in the 12-13\,$\mu$m region (see Fig.\,\ref{qeff_mir}), the contribution from the TiO$_{2}$ opacities will have a negligible impact in comparison to the MgAl$_{2}$O$_{4}$ opacities. Therefore, we decided to do the test by only adding spinel, in the same way as corundum, in the a-posteriori COMA radiative transfer computation, using the small-particle limit approximation. Refractive indices of spinel are adopted from \citet{zeidler11}, while the condensation temperature is fixed to $T_{\rm cond}^{\rm MgAl_{2}O_{4}}$=1150\,K (see e.g. \citealt{gail10}). Since aluminum dust is not included in the dynamical computation, we assume that spinel follows the same temperature-density structure as the gas. The degree of condensation ($\eta_{\rm MgAl_{2}O_{4}}$) is varied from 10\% to 100\% (with $\Delta\eta_{\rm MgAl_{2}O_{4}}$=10\%) to minimize the difference between the model and the visibilities for all baselines. We found the best agreement for a full condensation ($\eta_{\rm MgAl_{2}O_{4}}$=100\%) of Al condensing into spinel. 

\begin{figure}[tbp]
\begin{center}
\includegraphics[width=9.0cm]{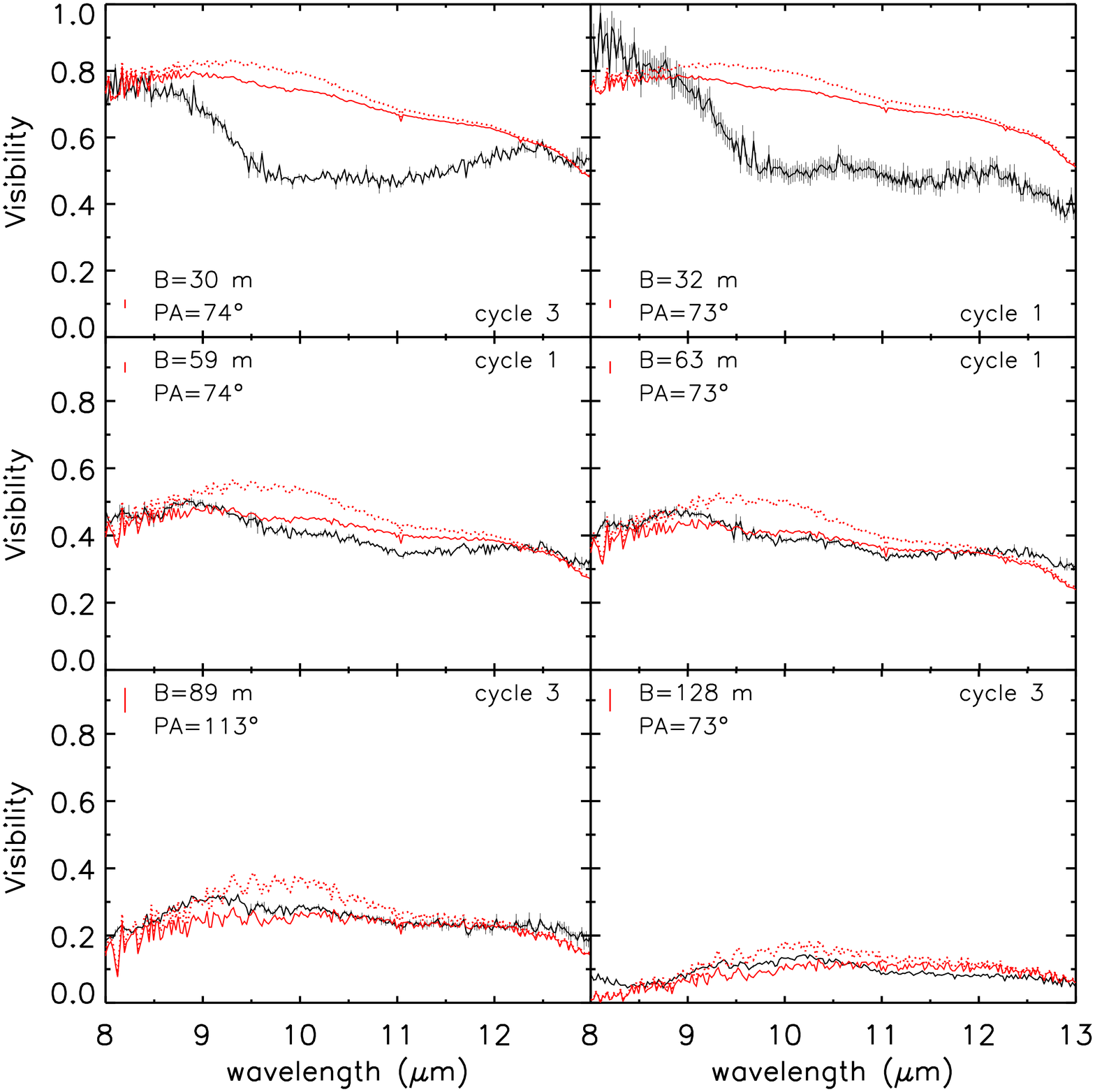}
\caption{Best-fitting synthetic visibility profiles with forsterite+aluminum of DMA~1 at phase $\phi_{\rm bol}$=0.60 and $\phi_{\rm bol}$=0.86 (red solid lines) together with the MIDI measurements of the star (black solid lines) for periods 2009 (cycle 1) and 2011 (cycle 3), respectively. Red bars, next to the legend, represent the average phase dispersion of the model. Red dotted lines represent the synthetic visibility profiles with aluminum dust only of DMA~1 at the same bolometric phases.} 
\label{COMPA_VIS_DMApureForst_500perceAl2O3_100perceMgAl2O4__alumina_only__onephase_MIDIdataRTVir}
\end{center}
\end{figure}

Figure\,\ref{COMPA_VIS_DMApureForst_500perceAl2O3_100perceMgAl2O4__alumina_only__onephase_MIDIdataRTVir} shows the comparison between the best-fitting synthetic visibility profiles with forsterite+aluminum of DMA~1 and the MIDI visibility measurements. Visibility measurements for baselines equal to and larger than 59\,m are reproduced well throughout the whole N-band range.\\

Obviously, a degree of condensation above 100\% (i.e. binding up more of an element in dust than what is available) for Al or other elements is not a realistic solution, no matter how good the fit of the synthetic visibilities is to the observed ones.
It should therefore be remembered that what is actually constrained by the fit is the column density of dust material contributing opacity to the relevant wavelength region. The dust column density is an integral containing the product of gas density times the degree of condensation. One possible solution to the problem is that the gas density in the circumstellar envelope in the close vicinity of \object{RT\,Vir} is actually higher than in our model. This would bring down the degrees of condensation by the corresponding factor, while keeping the same dust column density, in order to fit the observations in the N band. Considering that the mass-loss rates taken from the literature (see Sect.\,\ref{DMA_creation}) are uncertain by a factor of 3 or more, and are mean values spanning more than several hundred years of mass loss, while \object{RT\,Vir} shows clear cycle-to-cycle variability and long-term trends (see Sect.\,\ref{fc}), the current gas densities in the inner envelope are not well constrained by these values.

While a higher gas density (corresponding to a higher mass-loss rate) may help solve problems in the dust-dominated spectral regions, it could, on the other hand, also affect the molecule-dominated regions of the SED, in particular in the near-infrared. However, as illustrated by the current problems reproducing the CO${_2}$ feature (see Sect.\,\ref{dma_constraints}), questions remain regarding the validity of chemical equilibrium in the gas phase (e.g. \citealt{cherchneff06}) and other model assumptions which could affect the molecular spectra. Therefore a discussion of this problem is beyond the scope of the current paper.

\subsection{Identification of silicate grains in the 2 to 3 stellar radii region}
\label{add-confirm}

To confirm the presence of silicate dust in the two to three stellar radii region, the following test was performed. The COMA radiative transfer code was run only including aluminum dust (with no silicate dust included) for the 46 snapshots of DMA~1. Corundum and spinel dust grains were added in the same way as in the Sect.\,\ref{extra-dust}. Refractive indices of corundum and spinel were adopted from \citet{begemann97} and \citet{zeidler11}, respectively, while the condensation temperature was fixed to T=1400K for corundum and T=1150K for spinel. The only free parameter left is the degree of condensation of Al condensing into corundum and spinel. For that we varied the degree of condensation from 100\% to 1000\% (with steps of 100\%) to minimize the difference between the model and the visibilities for all baselines.

Figure\,\ref{COMPA_VIS_DMApureForst_500perceAl2O3_100perceMgAl2O4__alumina_only__onephase_MIDIdataRTVir} presents the best-fitting synthetic visibility profiles with aluminum dust only of the DMA~1 and the visibility measurements of \object{RT\,Vir}. We found the best agreement with the measurements for the same degree of condensation of Al ($\eta_{\rm Al_{2}O_{3}}$=500\% and $\eta_{\rm MgAl_{2}O_{4}}$=100\%) as the model that includes both forterite and aluminum dust (see Sect.\,\ref{extra-dust}).

Figure\,\ref{COMPA_VIS_DMApureForst_500perceAl2O3_100perceMgAl2O4__alumina_only__onephase_MIDIdataRTVir} shows a clear discrepancy (taking the average phase dispersion of the model into account) between the synthetic visibility profiles only with aluminum dust and the measurements around 9.8\,$\mu$m for the two 59 and 63\,m baselines measurements\footnote{Increasing the degree of condensation of Al condensing into corundum further makes the visibility profile of the model too steep in comparison with the measurements in the 10-13\,$\mu$m region without correcting the discrepancy around 9.8\,$\mu$m.}. This reveals a lack of opacity within the region around two to three stellar radii probed by these baselines at 9.8\,$\mu$m.
Among the data shown in Fig.\,\ref{qeff_mir}, only the silicate species show strong spectral features around 9.8\,$\mu$m. From Fig.\,\ref{qeff_mir}, it is difficult to distinguish between iron-rich and iron-free silicate dust in this wavelength region. However, detailed theoretical modeling \citep{woitke06} suggests that silicate grains forming within the region around two to three stellar radii have to be iron-free in order not to evaporate due to heating by the radiation field. Figure\,\ref{COMPA_VIS_DMApureForst_500perceAl2O3_100perceMgAl2O4__alumina_only__onephase_MIDIdataRTVir} shows that the model that includes both forsterite and aluminum dust (with the same degree of condensation of Al) can indeed reproduce the 9.8\,$\mu$m visibility measurements for the two 59 and 63\,m baselines, giving a clear confirmation for the presence of silicate in the two to three stellar radii spatial region.

The model including only aluminum dust (see Fig.\,\ref{COMPA_VIS_DMApureForst_500perceAl2O3_100perceMgAl2O4__alumina_only__onephase_MIDIdataRTVir}) also reveals that silicate dust has a very slight impact on spatial scales around two stellar radii (probed by the 89\,m baseline) and almost no influence on spatial scales around 1.5 stellar radii (probed by the 128\,m baseline), taking the average phase dispersion of the model into account. This is reasonable assuming that the heating by the radiation field within this region certainly leads to the evaporation of silicate dust but not to the evaporation of aluminum dust. 

This result not only confirms the presence of silicate dust in the region around two to three stellar radii, but also indicates that the column density of forsterite dust, provided by the DMA~1, is able to reproduce the spatial extent of the inner wind region within the 8-10\,$\mu$m spectral range.

\section{Exploring geometrical issues}
\label{discr}

Figure\,\ref{COMPA_VIS_DMApureForst_500perceAl2O3_100perceMgAl2O4__alumina_only__onephase_MIDIdataRTVir} reveals a lack of opacity of the model in comparison to the smallest 30\,m baseline visibility measurements. Such an opacity enhancement was also found around 9.8\,$\mu$m by the N-band geometrical modeling of \object{RT\,Vir} (see Sect.\,\ref{morphology} and upper-right plot of Fig.\,\ref{param_geo}). One possibility for overcoming this discrepancy is to artificially add an extra shell of dust material. 

\subsection{Addition of an extra shell of forsterite} 
\label{extra-shell}

We decided to include an extra shell of forsterite dust, in addition to our previous forsterite plus aluminum model (see Sect.\ref{extra-dust}), in order to reproduce the 30\,m baseline visibility data. To do that, we varied the position of the inner edge of the shell between four and ten stellar radii, corresponding to the spatial region probed by the 30\,m baseline. The abundance of forsterite is also varied under the assumption that 20, 40, 60, and 80\% Si has condensed into forsterite. The best fit is obtained for the location of the extra shell around four to six stellar radii away depending on the phase. The best forsterite abundance ratio is found to be $\eta_{\rm Mg_{2}SiO_{4}}$=60\%, while the abundance of aluminum grains are found to be $\eta_{\rm Al_{2}O_{3}}$=400\% and $\eta_{\rm MgAl_{2}O_{4}}$=80\%, keeping the same condensation temperatures of 1400\,K and 1150\,K, respectively. 

\begin{figure}[tbp]
\begin{center}
\includegraphics[width=9.0cm]{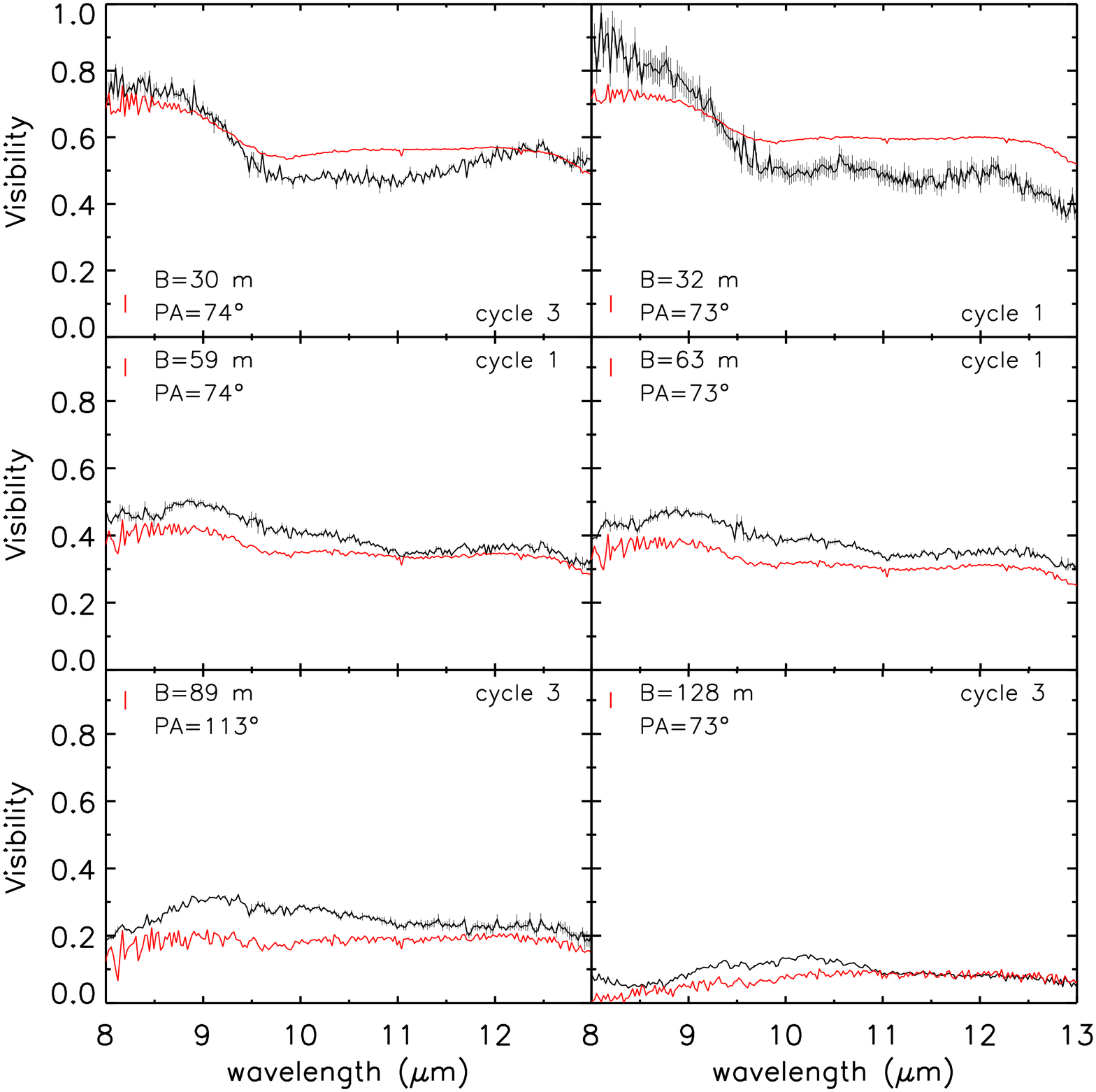}
\caption{Best-fitting synthetic visibility profiles with forsterite+aluminum and extra shell of forsterite of DMA~1 at phase $\phi_{\rm bol}$=0.60 and $\phi_{\rm bol}$=0.86 (red solid lines), together with the MIDI measurements of the star (black solid lines) for periods 2009 (cycle 1) and 2011 (cycle 3), respectively. Red bars, next to the legend, represent the average phase dispersion of the model.} 
\label{COMPA_VIS_DMApureForst_400perceAl2O3_80perce_MgAl2O4__ExtraLayer_forst_6Rstar_onephase_MIDIdataRTVir}
\end{center}
\end{figure}

\begin{figure}[tbp]
\begin{center}
\includegraphics[width=9.0cm]{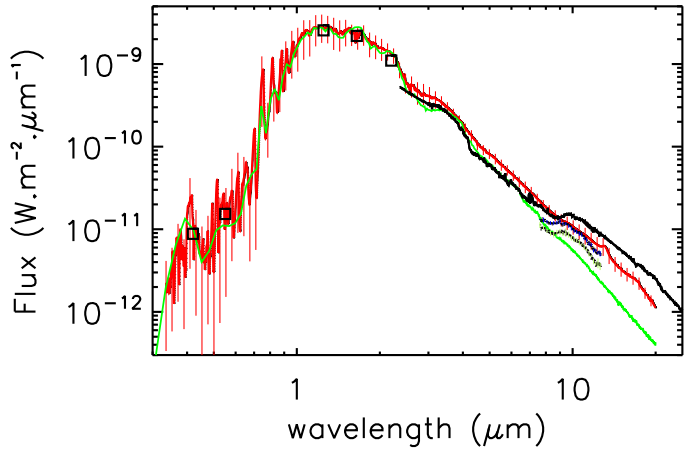}
\caption{Best-fitting synthetic spectrum with forsterite+aluminum and extra shell of forsterite of DMA~1 dispersed in phases (red solid line + bars), together with the broadband photometric data (black squares), ISO spectrum (black solid line), and MIDI spectra of \object{RT\,Vir} at phase $\phi_{\rm V}$=0.15 (Cycle 1) (blue error bars) and phase $\phi_{\rm V}$$\sim$0.0 (Cycle 3) (light green error bars). The best-fitting hydrostatic MARCS atmospheric spectrum (green solid line) is added too.}
\label{MARCS_HYDRODYN_PureMg2SiO4_opt4___400perceAl2O3_80perce_MgAl2O4__ExtraLayer_forst_6Rstar_onephase_phot_ISO}
\end{center}
\end{figure}

Results are shown in Figs.\,\ref{COMPA_VIS_DMApureForst_400perceAl2O3_80perce_MgAl2O4__ExtraLayer_forst_6Rstar_onephase_MIDIdataRTVir} and \ref{MARCS_HYDRODYN_PureMg2SiO4_opt4___400perceAl2O3_80perce_MgAl2O4__ExtraLayer_forst_6Rstar_onephase_phot_ISO}. Visibilities are now reproduced relatively well for all baselines, even if some discrepancies with the measurements still remain. The corresponding spectrum also shows a relatively good agreement with the measurements over the whole spectral range, especially revealing that the extra opacities added in the radiative transfer computation do not lead to any discrepancies in the near-infrared wavelength range.

\begin{figure}[tbp]
\begin{center}
\includegraphics[width=9.0cm]{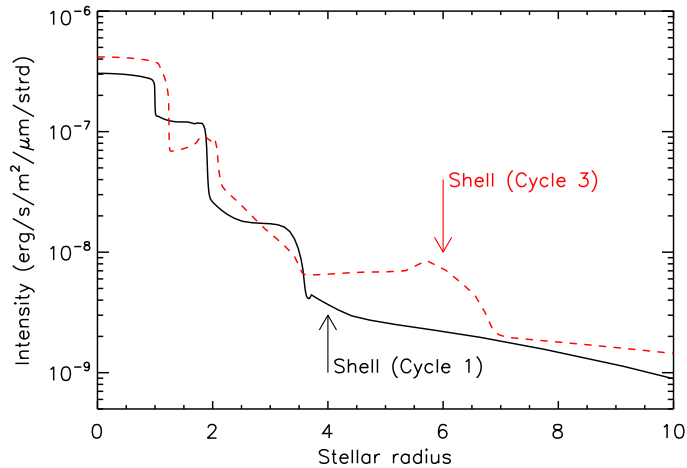}
\caption{Corresponding synthetic intensity profiles at 10\,$\mu$m, with forsterite+aluminum and the extra shell of forsterite, of DMA~1 at phases $\phi_{\rm bol}$=0.60 (black solid line) and $\phi_{\rm bol}$=0.86 (red dashed line).}
\label{Intensity_PureMg2SiO4_opt4___400perceAl2O3_80perce_MgAl2O4__ExtraLayer_forst_6Rstar}
\end{center}
\end{figure}

Figure.\,\ref{Intensity_PureMg2SiO4_opt4___400perceAl2O3_80perce_MgAl2O4__ExtraLayer_forst_6Rstar} presents the corresponding intensity profiles at 10\,$\mu$m, with forsterite plus aluminum dust and the extra shell of forsterite, of DMA~1 for phases $\phi_{\rm bol}$=0.60 and $\phi_{\rm bol}$=0.86. The extra shell is distinguishable from the horn-shaped profile around four stellar radii at phase $\phi_{\rm bol}$=0.60, and around six stellar radii at phase $\phi_{\rm bol}$=0.86. 
It is interesting to note that the position of the shell reproducing the cycle 1 interferometric measurements is around four stellar radii, while the position of the shell reproducing the cycle 3 interferometric measurements is around six stellar radii. This can be interpreted as shell having moved forward between the two cycles with a velocity of about 7.5\,km\,s$^{-1}$ (in good agreement with the time-average wind velocity of the best DMA~1, $<$u$>$=6.6 km\,s$^{-1}$).

\subsection{Possible explanations of the results}

Two scenarios, both related to the geometry of the circumstellar environment, can explain the need to add an extra dust shell to reproduce the interferometric measurements.

\begin{itemize}

\item \textit{Radial structure}\\
Because of the variability of the object, radial structure changes with time. The visual magnitudes of the star reveal an amplitude increase in its lightcurve over time. Moreover, the comparison between the MIDI and ISO/SWS spectra shows a cycle-to-cycle variability (see Sect.\,\ref{fc}). These changes can be related to a long-term variability of the star. Since we are only using a limited time-sequence of the DMA, we could be missing a period for which the density structure of the DMA gives better agreement with the interferometric measurements, without the need for extra silicate opacities.

\item \textit{Deviation from spherical symmetry}\\ 
Differential phase interferometric measurements reveal the presence of an asymmetry in the atmosphere of \object{RT\,Vir} (see Sect.\,\ref{morphology}). This is likely related to the presence of one or several clumpy structures in the close vicinity of the star, also suggested from water masers observations \citep{richards11}. Three-dimensional radiation hydrodynamics simulations of the convective interior and stellar atmosphere suggest there are non radial structures in the dust shells of AGB stars \citep{freytag08} that might explain these clumpy structures and could potentially affect the dynamical process of wind formation.

\end{itemize}

\section{Conclusion and perspectives}
\label{concl}

This work is a first attempt to understand the wind mechanism of M-type AGB objects by comparing photometric, spectrometric, and interferometric measurements with state-of-the-art, self-consistent, dust-driven wind models. \\

Comparison between spectro-interferometric data and a self-consistent dust-driven wind model reveals that silicate dust has to be present in the region between two to three stellar radii to reproduce the 59 and 63\,m baseline visibility measurements around 9.8\,$\mu$m. Detailed theoretical modeling \citep{woitke06} suggests that silicate grains forming on such spatial scales have to be iron-free in order not to evaporate owing to heating by the radiation field.

This result also indicates that the column density of forsterite dust, provided by the DMA, is able to reproduce the spatial extent of the inner wind region in the 8-10\,$\mu$m spectral range. This gives observational evidence, in addition to what is provided by \citet{norris12}, in favor of the theoretical predictions of \citet{hoefner08}, i.e., that winds can be driven by photon scattering on iron-free silicate grains formed in the close vicinity of an M-type AGB star.

However, the synthetic visibilities resulting from the model are above the measurements longward of 10\,$\mu$m, indicating that material strongly emitting around 11\,$\mu$m in the close stellar vicinity is missing in the radiative transfer computation. Ad-hoc addition of corundum and spinel dust in the a-posteriori radiative transfer can compensate for the lack of opacity, giving better agreement between the model and the 10-13\,$\mu$m interferometric measurements for baselines longer than 59\,m.\\

The necessity to increase the degree of condensation of Al material to a value about five times higher than the solar abundance, in order to obtain a good fit to the MIDI measurements, may indicate that the circumstellar gas density in the close vicinity of \object{RT\,Vir} is actually higher than in our model. The mass-loss rates used to constrain the dynamic model atmosphere of \object{RT\,Vir} are uncertain by a factor of 3, or more, and are mean values spanning more than several hundred years (e.g. \citealt{ramstedt08}). Considering that \object{RT\,Vir} shows clear cycle-to-cycle variability and long-term trends, the current gas densities in the inner envelope are not well constrained by these average values. On the other hand, a higher gas density could also affect the molecule-dominated regions of the SED, in particular in the near-infrared. However, as illustrated by the inability of the model to reproduce the CO${_2}$ feature, questions remain regarding the validity of chemical equilibrium in the gas phase and other model assumptions that could affect the molecular spectra.

The need for an extra silicate dust shell to reach the right level of the 30\,m baseline visibility measurements is still puzzling, but can be related to the geometry of the circumstellar environment in various ways. Among them, the long-term variability of the star and the presence of clumpy structures in the stellar vicinity are the most likely ones. To overcome these problems, a larger statistical study and progress in advanced self-consistent 3D modeling are required.\\

\begin{acknowledgements}

This work has been funded by the Swedish Research Council (Vetenskapsrådet). SR acknowledges support by the Deutsche Forschungsgemeinschaft (DFG) through the Emmy Noether Research grant VL 61/3-1. DK acknowledges support from the Austrian Science Fund FWF under project number AP2300621. BA acknowledges support from Austrian Science Fund (FWF) Projects AP23006 \& AP23586 and from contract ASI-INAF I/009/10/0. We would also like to thank the referee for valuable comments.

\end{acknowledgements}

%
%

%
%

\end{document}